\documentclass{iaus}
\usepackage{graphicx,natbib}

\title[Testing Convection Using Eclipsing Binaries]
{Testing Convection In Stellar Models Using Detached Eclipsing Binaries}

\author[Southworth \& Bruntt]{John Southworth$^1$ \and Hans Bruntt$^2$ }

\affiliation{$^1$\,Department of Physics, University of Warwick, Coventry, CV4 7AL, UK \break
             email: jkt@astro.keele.ac.uk\\[\affilskip]
             $^2$\,School of Physics A28, University of Sydney, 2006 NSW, Australia}

\pubyear{2007}
\volume{239}  
\pagerange{--}
\date{???? and in revised form ????}
\jname{Convection in Astrophysics}
\editors{F. Kupka, I.W. Roxburgh \& K.L. Chan, eds.}

\begin{document}

\maketitle

\begin{abstract}
The fundamental properties of detached eclipsing binary stars can be measured very accurately, which could make them important objects for constraining the treatment of convection in theoretical stellar models. However, only four or five pieces of information can be found for the average system, which is not enough. We discuss studies of more interesting and useful objects: eclipsing binaries in clusters and eclipsing binaries with pulsating components.
\end{abstract}

\firstsection 

\section{The problem}

Theoretical models of stellar evolution are of fundamental importance to stellar and galactic astrophysics because they are almost the only way of deriving the age, internal structure and composition of stars and galaxies from basic observational data (e.g.\ Meynet et al.\ 2005; Nordstr\"om et al.\ 2004). Unfortunately, their predictive power is severely limited by the parametric treatment of convective overshooting ($\alpha_{\rm OV}$), the efficiency of energy transport in turbulent matter ($\alpha_{\rm MLT}$), and chemical diffusion.

Changing $\alpha_{\rm OV}$ in theoretical models causes big changes in the predicted lifetimes, chemical yields and luminosities of massive stars (e.g.\ Maeder \& Meynet 1989). This in turn has a large effect on the predicted properties and formation rates of evolved objects such as core-collapse supernovae (Eldridge \& Tout 2004) and black holes. Changes in $\alpha_{\rm MLT}$ can affect the derived ages of globular clusters by up to 2\,Gyr (Chaboyer 1995), whilst reasonable changes in the amount of convective core overshooting can alter the derived ages of open clusters by a factor of two (Bragaglia \& Tosi 2006).

The predictive power of theoretical models must be improved by observationally constraining their treatment of convection. Detached eclipsing binaries (dEBs) have a lot of potential for doing this. The study of dEBs allows us to derive the masses and radii of two stars, of the same age and chemical composition, to accuracies of better than 1\% (Andersen 1991; Southworth et al.\ 2005b). Effective temperatures (and so luminosities and distances) can be found accurately from photometric calibrations or from modelling their spectra (Southworth et al.\ 2005ac). Such results have previously been used for calibrating theoretical models (e.g.\ Pols et al.\ 1997; VandenBerg et al.\ 2006) and then checking their predictions. However, previous studies (Andersen et al.\ 1990; Ludwig \& Salaris 1999; Ribas et al.\ 2000) have not been able to definitively pin down the values of $\alpha_{\rm OV}$ and $\alpha_{\rm MLT}$ or their dependence on mass and composition. Further information is required for each dEB before strong constraints can be applied to models.

For a dEB, the radii and effective temperatures are known for two specific masses. When comparing to theory, two of these four datapoints (usually the radii) are required to deduce the age and metal abundance of the system. The remaining two are generally not accurate enough to do anything more than indicate the helium abundance or, if you assume a normal helium abundance, perhaps the extent of convective overshooting. If you want to derive accurate constraints on core overshooting or mixing length, further information is needed. In many cases the surface metal abundance is observable, but usually isn't studied. Additional constraints are needed for each dEB.

\section{Solution 1: eclipsing binaries in open clusters}

\begin{figure}
\includegraphics[width=\textwidth,angle=0]{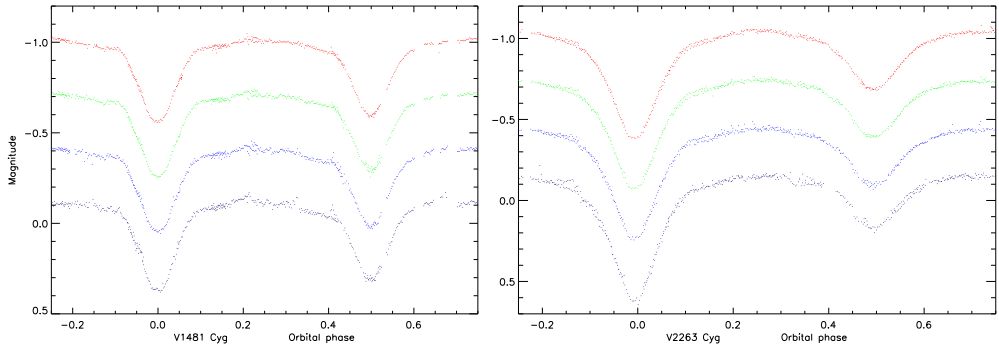}
\caption{\label{fig1} Preliminary (no debiassing or flat-fielding)
Str\"omgren $uvby$ light curves of the eclipsing binaries V1481\,Cyg and V2263\,Cyg in
NGC\,7128. V2263\,Cyg (right) is semi-detached so cannot be used to
constrain single-star evolutionary theory but will still be able to
provide an accurate distance to the cluster. $y$ is at the top and $u$ at the bottom.}
\end{figure}

The study of eclipsing binaries in stellar clusters allows more constraints to be placed on theoretical models (e.g.\ Thompson et al.\ 2000; Lebreton et al.\ 2001). In this case, the predictions of theoretical models must be able to match both the accurately-known properties of the dEB and the radiative properties of every other star in the cluster for the same age and chemical composition, resulting in more severe constraints on the theoretical description of convection. We are currently undertaking a research project to study a sample of dEBs in open clusters with a range of ages and chemical compositions.

Our first results (Southworth et al.\ 2004ad) concentrated on V615\,Per and V618\,Per, both members of the young open cluster NGC\,869. In conjunction with our analysis of V621\,Per (Southworth et al.\ 2004c) this has shown that the study of several dEBs in open cluster has great potential as a test of model predictions as it allows the unique possibility of placing four or more stars of the same age and chemical composition in one mass--radius or mass--temperature plot. Conversely, other results for dEBs in more sparse clusters (V453\,Cyg in NGC\,6871, Southworth et al.\ 2004b; DW\,Car in Cr\,228, Southworth \& Clausen 2006) have been less useful because the presence of other cluster stars has added little to the analysis. We have therefore obtained, in July 2006, 12 nights of wide-field imaging observations of the young open cluster NGC\,7128, which is known to contain at least six eclipsing binaries. Preliminary results are shown in Fig.~\ref{fig1}.

\section{Solution 2: eclipsing binaries with pulsating components}

The study of pulsating stars is also an excellent way in which to constrain the treatment of convection in theoretical stellar models. Such an analysis is helped by studying stars with accurately known properties, so we are currently conducting an observing program to obtain high-quality light curves of bright dEBs which contain pulsating components. Our photometry comes from the WIRE satellite, which typically observes each target for several weeks (for 30\% of each 90\,min Earth-orbit) with a scatter of 1--3\,$m$mag.

\begin{figure} \includegraphics[width=\textwidth,angle=0]{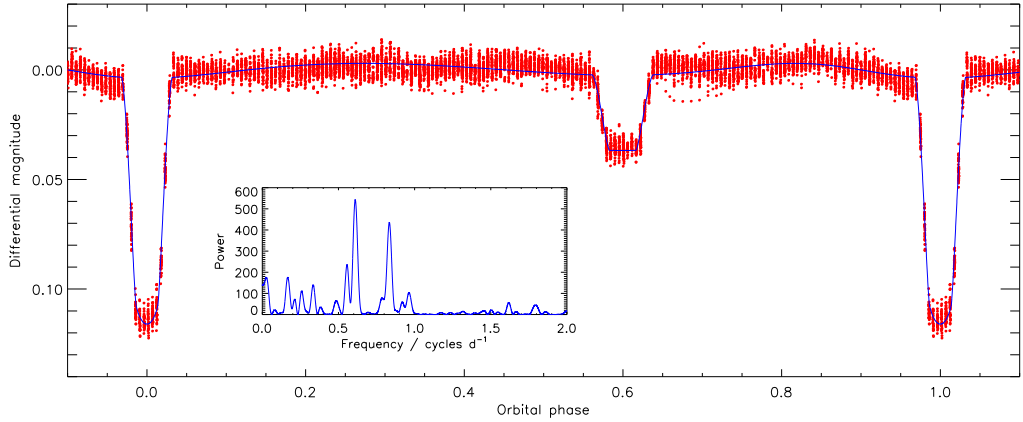}
\caption{\label{fig2} Preliminary fit to the the light curve of AR\,Cas
(main panel) with a periodogram of the residuals (inset panel) showing
strong periodicity at the rotational period of the primary star (1.6\,days;
amplitude 2\,$m$mag) and at several other periods.} \end{figure}

Our first target was the A-type star $\psi$\,Cen (Bruntt et al.\ 2006) which we discovered to be a long-period dEB. We have measured the relative radii of the stars (expressed as a fraction of the orbital separation) to unprecedented accuracies of 0.1\% and 0.2\%, and have found $g$-mode pulsations with frequencies of 1.996 and 5.127 cycles day$^{-1}$ and amplitudes of 0.23 and 0.18 $m$mag in the residuals of this solution. We are currently obtaining the radial velocity observations needed to measure masses and so provide strong constraints on theoretical models. Our second target is AR\,Cas (Fig.~\ref{fig2}), which has shallow total eclipses on a 6.0\,day period and at least six photometric frequencies.

\end{document}